# A Tableau Methodology for Deontic Conditional Logics

Alberto Artosi[1] and Guido Governatori[2]

[1] Department of Philosophy, University of Bologna, via Zamboni 38, 40126,
+39-51-258333, email: artosi@cirfid.unibo.it
[2] Department of Computing, Imperial College, 180 Queen's Gate, London, SW7 2BZ,
+44-(0)171-59 48210, email: gg6@doc.ic.ac.uk

**Abstract.** In this paper we present a theorem proving methodology for a restricted but significant fragment of the conditional language made up of (boolean combinations of) conditional statements with unnested antecedents. The method is based on the possible world semantics for conditional logics. The label formalism introduced in [AG94,ABGR96] to account for the semantics of normal modal logics is easily adapted to the semantics of conditional logics by simply indexing labels with formulas. The inference rules are provided by the propositional system $KE^+$ — a tableau-like analytic proof system devised to be used both as a refutation and a direct method of proof — enlarged with suitable elimination rules for the conditional connective. The theorem proving methodology we are going to present can be viewed as a first step towards developing an appropriate algorithmic framework for several conditional logics for (defeasible) conditional obligation.

## 1 Introduction

If you promised to marry Suzy Mae, then you ought to marry her, but there is no obligation to marry her if you promised under menace of Suzy Mae's brothers. Your obligation is, so to speak, defeated or overridden by the additional circumstance of your having promised under the menace of Suzy Mae's brothers. This amounts to saying that deontic conditionals are very often *defeasible*. The distinctive mark of a defeasible conditional is that it fails both MP and monotony (*alias* strengthening antecedents). For example, from the fact that $A$ is the case, and that it is obligatory to do $B$ when $A$ is the case, it follows that it is obligatory to do $B$ only if it further assumed that the circumstances are unexceptional, so that your having promised under coercition does not allow us to detach that you ought to marry Suzy Mae in spite of your having promised to do so. Since Lewis's classic account of counterfactuals in [Lew86], conditional logic ($CL$) and its possible worlds semantics have been advocated as an appropriate tool for the analysis of deontic conditionals (see e.g. [Mot73,Che74]. This approach has been anticipated by [Han69,vFr72]). Lewis's semantics makes invalid the monotony, and in some versions also the MP, formula. For example, the failure of your obligation to follow from your promise, given in addition that you promised under coercition, may be explained in term of the "distance" between the (best) possible worlds in which your promise commits you to marry Suzy Mae and the (best of the much more distant) worlds where you promised under coercition.

Unfortunately, $CL$s are not particularly well suited for investigating non-monotonic forms of inference in proof-theoretic terms. In fact, in contrast with the striking development of $CL$'s semantic setting, its inferential structure has remained largely unexplored (notable exceptions are [Tho70] and [Swa83]). Most importantly, it has not been sufficiently explored to provide automated deduction methods for effectively computing non-monotonic inference relations. We know only two attempts in this direction: Groeneboer and Delgrande's [GD88] and Lamarre's [Lam92] tableau-based theorem provers for some normal $CL$s. Our purpose in this paper is to outline a tableau methodology for a special fragment of $CL$'s language. This language is poorer than that used in Lewis's $CL$s: iterated conditional connectives are allowed only in the consequent of conditional formulas. This restriction arises naturally out of the deontic, and in general defeasible logic, literature. Almost all logics for defeasible conditionality use only the "flat" (i.e. non nested) part of $CL$, (see e.g., [Alc93,Bel90,Bel91,Del87,Del88,Han69,KLM90,Lam91,Leh89]), and some admit iteration of the conditional connective only in the consequent of conditional formulas (this is the case with [CT92]'s four-valued conditional logic for knowledge update and [GMS96]'s treatment of hypothetical logic programming). As we shall see, this restriction has its pros and cons. They will be discussed in the final section (together with some open questions). In Section 2, we expound Lewis-type semantics for deontic $CL$s. In Section 3 we describe a label formalism adequate to represent $CL$'s semantics. In Section 4 we present the full proof method and discuss some main problems.

## 2 Lewis's $CL$s for Conditional Obligation

We begin with the language and the standard possible worlds model theory for $CL$. Let $L_>$ be the language obtained from that of classical propositional logic by adding the symbol $>$ for the conditional connective. The set of (well-formed) formulas of $L_>$ is defined in the usual way. If $A$ is a formula of $L_>$, then we write $A \in L_>$. The formal semantic interpretation of $L_>$ is in terms of *system of spheres* ($SOS$) models. An $SOS$ model [Lew86,Nut80] is a triple $\langle \mathcal{W}, S, v \rangle$ where $\mathcal{W}$ is a nonempty set (of possible worlds); $v$ is a valuation assigning to each $u$ in $\mathcal{W}$ and $A \in L_>$ an element from the set $\{T, F\}$ (if $v(A, u) = T$, then we call $u$ an $A$-world), and $S$ is a function which assigns to each $u$ in $\mathcal{W}$ a nested set $S(u)$ of subsets of $\mathcal{W}$ closed under unions and nonempty intersections (the *system of spheres of similarity around $u$*). The intuitive idea is that, for a world $u$, $S(u)$ represents different degrees of similarity to $u$: if $s \in S(u)$, then any world in $s$ is considered to be closer or more similar to $u$ than any world not in $s$. The truth conditions for classical formulas in an $SOS$-model are as usual. Let us call a sphere $s \in S(u)$ containing at least an $A$-world an *Asphere*. The rule for evaluating conditional formulas $A > B$ is thus given by: $A > B$ is true at a world $u$ iff either 1) there is no $A$-world in any sphere $s \in S(u)$ ($A > B$ is vacuously true), or 2) $B$ holds at every $A$-world in the smallest $A$-sphere in $S(u)$ (i.e. in every $A$-world closest to $u$). The notion of validity of a formula is defined as usual, as truth in all worlds in all $SOS$-models.

In the deontic interpretation we assume that members of $\mathcal{W}$ are ordered from the standpoint of any $u$ according to some comparative standard of (moral or legal) evaluation. For a world $u$, $s \in S(u)$ represents now a set of worlds evaluable from $u$ such that any world in $s$ is considered to be better that any world not in $s$ (in the ordering) from the standpoint of $u$. Under this interpretation, the conditional formula $A > B$ is to be interpreted as "If (or given that) $A$, it ought to be that $B$" with (*mutatis mutandis*) the same truth-condition as before. Notice that clause 2) of the above definition assumes that there is a *smallest* $A$-sphere in $S(u)$ (i.e., one that is included in all the others) containing the $A$-worlds *closest* to — or, in the deontic interpretation, *the best* $A$-worlds from the standpoint of — $u$. Hence no infinite sequence of better and better worlds is allowed. As it is well known this assumption (" limit assumption") is highly questionable (though it does not affect validity). Since our emphasis is not on conceptual issues, but rather on getting an effective methodology for theorem proving in conditional logics, we shall avoid here to discuss whether infinite sequences are a "serious possibility".

The class of all $SOS$-models characterizes Lewis's weakest system of $CL$ (the system $V$). In the deontic interpretation one may wish to assume one or more of the following constraints on $SOS$-models. Let us call $S$ 1) *normal* iff $\bigcup S(u) \neq \emptyset$ for all $u \in \mathcal{W}$ (sphere systems are not empty, i.e. there is at least one world which is evaluable from $u$); 2) *universal* iff $\bigcup S(u) = \mathcal{W}$ for all $u \in \mathcal{W}$ (all worlds are evaluable from $u$); 3) *absolute* iff $S(u) = S(v)$ for all $u \in \mathcal{W}$ (the preference ordering is the same for all worlds). An $SOS$-model will be called *normal, universal, absolute* according to whether $S$ is normal, universal, absolute. Normal, universal and absolute $SOS$-models characterize respectively Lewis's systems $VN$, $VTU$ and $VA$. Absolute normal and absolute universal $SOS$-models characterize respectively Lewis's systems $VNA$ and $VTA$. All these systems (including $V$) are proposed by Lewis as suitable for a logic of conditional obligation. (For a more refined treatment and comparison with traditional dyadic deontic logic see [Lew74]). It should be noticed that all the kinds of $SOS$-models mentioned above fail to validate both monotony and MP for $>$ (respectively, $(A > B) \to ((A \wedge C) > B)$ and $(A > B) \to (A \to B)$). Let us recall that the latter is validated by (weakly) centered $SOS$-models. Lewis's reason for rejecting such models as suitable for deontic interpretation is that on their assumption each world would be, from its own standpoint, the best of all (or at least one of the best) possible worlds.

Our next step will be to show how the $SOS$ semantic model construction just presented can be mirrored in an adequate label formalism.

## 3 Label formalism and conditional semantics

In [AG94,ABGR96] we presented a labelled proof system for normal modal logics, called $KEM$, which seems to enjoy most of the features a suitable proof search system for modal (and in general non-classical) logics should have. In particular, as we have argued elsewhere, it appears to be flexible enough to be extended to cover the full range of non-classical logics which are extension of (or logically similar to) modal logic — indeed flexible enough to be adapted to any setting having a Kripke-model based semantics (see e.g. [AGS96]). This is largely due to the particular label formalism it uses to generate and check models. In this section we show how this formalism can be extended, almost without modification, to handle the semantics of $CL$.

As we have said the $KEM$ approach requires us to work with "world" labels (thereafter labels). Let $\Phi_C = \{w_1, w_2, \dots\}$ and $\Phi_V = \{W_1, W_2, \dots\}$ be two (non empty) sets respectively of constant and variable world-symbols. A *label* in the sense of [AG94,ABGR96] is an element of the set $\Im$ defined as follows:

$\Im = \bigcup_{1 \leq i} \Im_i$ where $\Im_i$ is:
$\Im_1 = \Phi_C \cup \Phi_V$;
$\Im_2 = \Im_1 \times \Phi_C$;
$\Im_{n+1} = \Im_1 \times \Im_n$.

Thus, a label is either a constant or a variable world-symbol (in which case we speak of *atomic* labels), or a "structured" sequence $(k', k)$ of world-symbols where (i) $k' \in \Phi_C \cup \Phi_V$, and (ii) $k \in \Phi_C$ or $k = (m', m)$ where $(m', m)$ is a label. From now on we shall use $i, j, k, \dots$ to denote arbitrary labels.

As far as Kripke models for normal modal logics are concerned, we may think of constant and variable world-symbols as denoting respectively worlds and sets of worlds (*any* world) in a Kripke model. A label of the form $(k', k)$ may be viewed as representing a "world-path". For instance, the label $(W_1, w_1)$ represents a path from $w_1$ to the set $W_1$ of worlds accessible from $w_1$; $(w_2, (W_1, w_1))$ represents a path which takes us to a world $w_2$ accessible by any world accessible from $w_1$ (i.e., accessible by the sub-path $(W_1, w_1)$) according to the appropriate accessibility relation. Thus a label of the form $(k', k)$ is "structurally" designed to convey information about access between the worlds in it.

In passing from Kripke models for modal logics to $SOS$-models for $CL$s, the format of the labels is left unchanged. The only modification is that they are now indexed by formulas (called *label formulas*). For a formula $A$, $w_1^A$ and $W_1^A$ will stand respectively for an $A$-world (in some $A$-sphere) and a set of such worlds (any $A$-world in some $A$-sphere). The interpretation of labels of the form $(k', k)$ varies accordingly. For example, the labels $(w_1^A, w_1)$ and $(W_1^A, w_1)$ can be viewed as representing respectively an $A$-world and any $A$-world in the smallest $A$-sphere around $w_1$. Similarly, the label $(w_2^A, (W_1^{A \vee C}, w_1))$ represents an $A$-world in the smallest $A$-sphere with respect to whatever world in the smallest $A \vee C$-sphere around $w_1$. Such a reading is motivated by the general idea [Che75] that the conditional $A > B$ can be regarded as a unary operator dependent on $B$ (i.e.

$[A >]B)$, from which it follows that whenever $A > B$ is true at $w_1$, $B$ is true at all $A$-worlds in the smallest $A$-sphere around $w_1$; and whenever $A > B$ is false at $w_1$, there is some $A$-world $w_2^A$ in the smallest $A$-sphere around $w_1$ at which $B$ is false.

In $KEM$ original approach we attached labels to *signed* formulas (i.e. formulas of the modal language prefixed with a "$T$" or "$F$") to yield pairs of the form $X, i$, with $X$ a signed formula and $i$ a label, called *labelled signed formulas* (*LS-formulas*). Intuitively, an *LS*-formula, $TA, i$ ($FA, i$) is intended to mean: $A$ is true (false) at the (last) world(s) (on the path) $i$; for instance, $TA \to B$, $(W_1, w_1)$ means that $A \to B$ is true at all the worlds (any world) accessible from $w_1$. The key feature of $KEM$ approach is that in the course of proof search labels are manipulated in a way closely related to the semantics of modal operators and "matched" using a specialized unification algorithm. That two labels $i$ and $k$ are unifiable means, intuitively, that any world which one could get by the path $i$ could be reached by the path $k$ and vice versa (equivalently, that the sets of worlds they "denote" have a non-null intersection). For example, $(w_3, (W_1, w_1))$ and $(W_3, (w_2, w_1))$ are unifiable (by simultaneously linking $W_3$ to $w_3$ and $W_1$ to $w_2$); thus they virtually represent the same path (since $w_3$ is a world in $W_3$ and $w_2$ is a world in $W_1$). *LS*-formulas whose labels are unifiable turn out to be true (false) at the same world(s) relative to the accessibility relation that holds in the appropriate class of models. In particular two complementary *LS*-formulas $TA, i, FA, k$ ($FA, i, TA, k$) whose labels are unifiable stand for formulas which are contradictory "in the same world".

This extends immediately to $CL$'s semantics. For instance, a *LS*-formula $TC, (W_1^{A \vee B}, w_1)$ means that $C$ is true at all $A \vee B$-worlds (any $A \vee B$-world) in the smallest $A \vee B$-sphere around $w_1$. That two labels, e.g., $(W_1^{A \vee B}, w_1)$ and $(w_3^{A \vee B}, w_1)$, are unifiable will mean that $w_3^{A \vee B}$ is an $A \vee B$-world in the smallest $A \vee B$-sphere around $w_1$. Thus the pair of *LS*-formulas $TC, (W_1^{A \vee B}, w_1)$ and $FC, (w_3^{A \vee B}, w_1)$ expresses a contradiction in the same $A \vee B$-world in the smallest $A \vee B$-sphere around $w_1$.

To distinguish elements of labels we shall use the following terminology and notation: given a label $i = (j, k)$, $j$ is said to be the *head* of $i$, denoted by $h(i)$, and $k$ the *body* of $i$, denoted by $b(i)$. By definition both $h(i)$ and $b(i)$ are labels, so that we can apply recursively such notions. For example, $h(b(i))$ and $b(b(i))$ will denote respectively the head and the body of the label $b(i)$. We shall also call each $b(i), b(b(i)), \ldots$ a *segment* of $i$ and denote it by $s(i)$. Intuitively, $s(i)$ stands for a sub-path of $i$. The length of a label $i$, $l(i)$, is the number of world symbols it is made of. We use $s^n(i)$ to denote the segment of $i$ whose length is $n$. Similarly we shall use $h^n(i)$ to name the $n$-th world-symbol in the label (counting from the right). The notion of the *countersegment*-n of a given label $i$ is defined as follows:

$$c^n(i) = h(i) \times (\cdots \times (h^k(i) \times (\cdots \times (h^{n+1}(i), w_0)))) \ (n < k < l(i))$$

where $w_0$ is a "dummy" label, i.e. a label not appearing in $i$ (the context in which such a notion is applied will tell us what $w_0$ stands for).

### 3.1 Label unification

As we have already said, the logic-dependent notion of label (or $\sigma$-) unification is at the heart of $KEM$ modal proof system. In this section we define a special notion of $\sigma$-unification for $CL$ (let us denote it by $\sigma_>^{CL}$). Let us first recall that a label indexed with a formula $Y$ (of $L_>$) is meant to denote either a $Y$-world or a set of $Y$-worlds in some $Y$-sphere. Thus $(W_1^\top, w_1)$, denotes the set of worlds (in the deontic interpretation) evaluable from $w_1$. Notice that $W^\top = \mathcal{W}$ since $\top$ holds in all worlds. We base our notion of $\sigma_>^{CL}$-unification on the $\sigma_L$-unification (see [AG94,AGS96,ABGR96,Gov97] for a definition) for the corresponding "outer" (normal) modal logic $L$ (see [Lew86], Section 6.3, for a list of $CL$s and the corresponding outer modal logics). In general, each $\sigma_L$ has several turning points, i.e., pairs of segments that should be matched (unified). In order to obtain $\sigma_>^{CL}$ we impose some constraints on the formulas indexing such turning points. In what follows we deal only with a very basic condition (for certain $CL$s the notion of $\sigma_>$-unification might require more cumbersome conditions).

$$(i^Y, j^Z)\sigma_> = (i, j)\sigma$$

and for each turning point $(i'^{Y'}, j'^{Z'})$ one of the following conditions holds

- $Y' \equiv Z'$, or
- $Y' \equiv \top$ and $h(i) \in \Phi_V$ or
- $Z' \equiv \top$ and $h(j) \in \Phi_V$

We shall say that a label $i$ *extends* a label $k$ iff there exists an $s(i)$ such that either (i) $s(i) = k$ or (ii) $(s(i), k)\sigma_>$; and that $i$ *extends immediately* $k$ iff $i$ extends $k$ and $s(i) = b(i)$.

## 4 Tableaux for $CL$s

In this section we rely on the label formalism just presented, together with appropriate (tableau like) inference rules, to provide a theorem proving methodology for a restricted fragment $L_>^-$ of $L_>$ made up of (boolean combinations) of formulas of the form $A > B$ with no nested occurrences of $>$ in the antecedent.

### 4.1 Inference rules

According to the interpretation for labels and the conditional connective expounded in the previous section the inference rules for $>$ would turn out to be as follows:
$$\frac{TA > B, i}{TB, (W_n^A, i)} \quad \frac{FA > B, i}{FB, (w_n^A, i)}.$$

In this format, they closely resemble the $KEM$ inference rules for the usual (unary) modalities [AG94,ABGR96]. However, as they stand, they are misleadingly straightforward. In fact, using these rules causes some information to be lost in the resulting proof tree (this turns out to be immediately evident if we try to construct a $KEM$ tree for the V axiom $((A > B) \wedge (B > A)) \rightarrow ((A > C) \equiv (B > C)))$. To prevent this, we have to draw conclusions from both the declarative and the labelled part of $LS$-formulas. To this end we introduce the following notation:

$$SA : X, i^Y \iff \begin{cases} X = SA \text{ or} \\ Y = A \text{ and } S = T \end{cases}$$

where $S \in \{T, F\}$, $A, Y \in L_>^-$, $X$ is a signed formula, and $i \in \Im$.

We are now ready to define the full set of inference rules for $CL$ as follows. We assume familiarity with Smullyan [Smu68] uniform notation; we write $(i, j)\sigma$ to denote both that $i$ and $j$ are $\sigma$-unifiable and the result of their $\sigma$-unification, and $X^C$ to denote the conjugate of a signed formula $X$.

$$\frac{\alpha : X, k^Y}{\alpha_i, k^Y}[i = 1, 2] \tag{$\alpha$}$$

$$\frac{\beta : X, k^Y \quad \beta_{3-i}^C : X', j^{Y'}}{\beta_i, (k, j)\sigma_L}[i = 1, 2] \tag{$\beta$}$$

These are exactly the $\alpha$ and $\beta$ rules of the original $KEM$ method in a slightly modified version (the $\alpha$ rules are just the usual linear branch-expansion rules of the tableau methods, whereas the $\beta$ rules correspond to such common natural inference patterns as modus ponens, modus tollens, disjunctive syllogism, etc.).

$$\frac{TA > B : X, i^Y}{TB, (W_n^A, i^Y)} \tag{$T > 1$}$$

$$\frac{TA > B : X, i^Y \quad TA\,(FB) : X', i'^A}{TB\,(FA), (c^{l(i')-1}(i'))^A} \tag{$T > 2$}$$

if $i'$ extends immediately $i$ (i.e. $(b(i'), i)\sigma_L$) and $w_0 = (b(i'), i)\sigma_L$.

The presence of two rules for $T >$ is due to the fact that true conditional formulas behave both as $\alpha$- and $\beta$-formulas (according to Smullyan classification). Note, however, that $T > 2$ may result problematic for certain $CL$s.

$$\frac{FA > B, i}{FB, (w_n^A, i)} \tag{$F >$}$$

$$\frac{\quad}{X,i \quad X^C,i}[i \text{ unrestricted}] \tag{PB}$$

$$\frac{\begin{array}{c} X:Y,i^{Y'} \\ X^C:Z,k^{Z'} \end{array}}{\times}[(i,k)\sigma_L] \tag{PNC}$$

PB (the "Principle of Bivalence") is exactly the "cut" rule of the original $KEM$ method (it represents the semantic counterpart of the cut rule of the sequent calculus), whereas PNC (the "Principle of Non-Contradiction") is the modified version of the original $KEM$ (i.e., tableau) branch-closure rule. Notice that, in contrast with normal modal logics, in $CL$'s setting it may occur a contradiction of the form $FA,(w_2^A,w_1)$, since what is said by this $LS$-formula is that $A$ is false at an $A$-world (a world at which $A$ is true) in the smallest $A$-sphere around $w_1$.

### 4.2 Problems

It is easy to see, from the above definition of $\sigma_>$- unification and the form of the inference rules, what problems arise for a tableau system for $CL$s. We shall examine some of them with the help of examples. Let us first consider the following formula of $L_>^-$.

$$((A \vee \neg A) > B) \to (C > B) \tag{1}$$

This formula is valid. Let us suppose that it is not. Then the antecedent is true and the consequent false. But the former is equivalent to $\top > B$ (which means that $B$ holds in all the $\top$-worlds in some $\top$-sphere), whereas according to the consequent there is a $C$-world in the smallest $C$-sphere where $B$ is false. Thus, since a $C$-world is also a $\top$-world, there is a $\top$-world where $B$ is at the same time true and false. Let us see the $KEM$-tree for this formula.

1. $F((A \vee \neg A) > B) \to (C > B)$      $w_1$
2. $T(A \vee \neg A) > B$      $w_1$
3. $FC > B$      $w_1$
4. $TB$      $(W_1^{A \vee \neg A}, w_1)$
5. $FB$      $(w_2^C, w_1)$

The steps from 1 to 5 are straightforward. At this point we have two complementary formulas (4 and 5), so we have to check whether they are also $\sigma_>$-complementary (i.e complementary under the $\sigma_>$-unification of their labels). In this case we have to go on with the proof, because the label formula $A \vee \neg A$ has not yet been analysed. Let $\mathcal{D}$ stand for the first 5 steps of the proof. Thus we have:

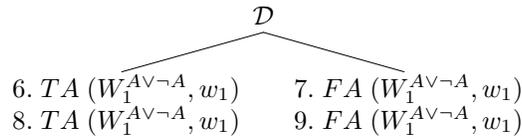

6. $TA\ (W_1^{A \vee \neg A}, w_1)$      7. $FA\ (W_1^{A \vee \neg A}, w_1)$
8. $TA\ (W_1^{A \vee \neg A}, w_1)$      9. $FA\ (W_1^{A \vee \neg A}, w_1)$

It is immediate to see that both branches are open, but a closer inspection of the tree reveals that it satisfies the condition of Theorem 1 below, thus $A \vee \neg A \equiv \top$, from which it follows that $(W_1^{A \vee \neg A}, w_1)$ is equivalent to $(W_1^{\top}, w_1)$. This let it to $\sigma_>$-unify with $(w_2^C, w_1)$ thus closing the tree. Alternatively we could open a tree for the negation of all non-atomic label formulas.

Let us now consider the formula

$$(A \wedge \neg A) > B \qquad (2)$$

which holds in all the $V$-logics, with the corresponding tree

$$
\begin{array}{ll}
1.\ F(A \wedge \neg A) > B & w_1 \\
2.\ FB & (w_2^{A \wedge \neg A}, w_1)
\end{array}
$$

What tree have we now to develop? That for $TA \wedge \neg A$ or that for $FA \wedge \neg A$? According to the above remarks, we would have to develop the tree for $FA \wedge \neg A$, but in the present case this will not led us to the the desired result, since all we learn from an open tree is that the negation of the formula is satisfiable. On the other hand, going on with the application of the inference rules leads to the following tree:

$$
\begin{array}{ll}
1.\ F(A \wedge \neg A) > B & w_1 \\
2.\ FB & (w_2^{A \wedge \neg A}, w_1) \\
3.\ TA & (w_2^{A \wedge \neg A}, w_1) \\
4.\ FA & (w_2^{A \wedge \neg A}, w_1) \\
5.\ \times &
\end{array}
$$

Another question concerns the treatment of equivalences. A common inference rule for $CL$s is

$$\frac{A \equiv B}{(A > C) \to (B > C)} \qquad (3)$$

from which it follows

$$((\neg A \vee B) > C) \to ((A \to B) > C) \qquad (4)$$

The easiest solution might seem to open an auxiliary tree each time we have to check whether two labels unify, which means that the label formulas turn out to be equivalent[1]. Indeed, this solution leads to a remarkable increase in the complexity of proof as the following tree shows.

$$
\begin{array}{ll}
1.\ F((\neg A \vee B) > C) \to ((A \to B) > C) & w_1 \\
2.\ T(\neg A \vee B) > C & w_1 \\
3.\ F(A \to B) > C & w_1 \\
4.\ TC & (W_1^{\neg A \vee B}, w_1) \\
5.\ FC & (w_2^{A \to B}, w_1)
\end{array}
$$

---
[1] See Section 3.1.

Notice that the tree closes if the labels of the complementary formulas $TC$ and $FC$ $\sigma_>$-unify, which happens iff $\neg A \vee B$ is equivalent to $A \to B$. Accordingly, we have to open a new tree for proving their equivalence:

$$F(\neg A \vee B) \equiv (A \to B)$$

| $T\neg A \vee B$ | $F\neg A \vee B$ |
|---|---|
| $FA \to B$ | $TA \to B$ |
| $TA$ | $TA$ |
| $FB$ | $FB$ |
| $TB$ | $TB$ |

### 4.3 The classical system $KE^+$

The shortcomings discussed in the previous section are avoided by appealing to the propositional proof system $KE^+$. $KE^+$ is based on D'Agostino and Mondadori's [DM94] $KE$, a tableau-like classical proof system which employs a mixture of tableau, natural deduction and structural rules (in practice, the $\alpha$-, $\beta$-, PB and PNC rules of Section 4.1 restricted to the propositional part). $KE^+$ uses the same rules but adopts a different proof search procedure which makes it completely analytical and able to detect whether 1) a formula is either a tautology, or a contradiction, or only a satisfiable one; and 2) a sub-formula of the formula to be proved is a tautology and to use this fact in the proof of the initial formula.

The $KE^+$ based method consists in verifying whether the truth of the conjugate of an immediate sub-formula of a $\beta$-formula implies the truth of the other immediate sub-formula. If it is so, then we have enough information to infer that the whole formula is provable. This result follows from the fact that in a given branch — the branch beginning with $\beta_i^C (i = 1, 2)$ — a formula leading to closure does not exist, i.e., there is no pair of complementary formulas. This is obtained by proving that the conjugate of the formula occurs in the branch, i.e., we have to see whether a signed formula appears twice in it, and that the two occurrences are derived from appropriate formulas. We now see how $KE^+$ does such a job.

### 4.4 Proof Search with $KE^+$

As said before, the key feature of $KE^+$ is that it can be used to determine tautologicity of sub-formulas in the course of a refutation proof. To this end we need the following definitions.

**Definition 1.** *Each formula depends on itself. A formula B depends on A either if it is obtained by an application of the $\alpha$-rule or it is obtained by an application of KE's rules on formulas depending on A. A formula C depends on A, B if it is obtained by an application of a $\beta$-rule with A, B as premises. The formulas obtained by an application of PB depend on the formula PB is applied to. If C depends on A, B then C depends on A and C depends on B.*

**Definition 2.** *An $\alpha$-formula is* analysed *in a branch when both $\alpha_1$ and $\alpha_2$ are in the branch. A $\beta$-formula is* analysed *in a branch when either 1) if $\beta_1^C$ is in the branch also $\beta_2$ is in the branch, or 2) if $\beta_2^C$ is in the branch also $\beta_1$ is in the branch. A $\beta$ formula will be called* fulfilled *in a branch if: 1) either $\beta_1$ or $\beta_2$ occurs in the branch provided they depend on $\beta$, or 2) either $\beta_1$ or $\beta_2$ is obtained from applying PB on $\beta$.*

**Definition 3.** *A branch is* E-completed *if all the formulas occurring in it are analysed. A branch is* completed *if it is E-completed and all the $\beta$-formulas occurring in it are fulfilled.*

**Definition 4.** *We shall call a branch a $\beta^C$-branch if its root is obtained by applying PB to a $\beta$-formula and it starts with $\beta_i^C$. Each branch generated by an application of PB to a formula occurring in a $\beta^C$-branch is a $\beta^C$-branch. Any branch which is not a $\beta^C$-branch and is obtained from an application of PB will be called a $\beta$-branch.*

**Definition 5.** *We shall call a branch a $\top$-branch if it contains only formulas which are equivalent to $\top$ and the formulas depending on them.*

The procedure starts with the formula to be proved; then

1. it selects a $\beta^C$-branch $\phi$ which is not yet completed and which is the $\beta^C$-branch with respect to the greatest number of formulas;
2. if $\phi$ is not $E$-completed, it expands $\phi$ by means of the $\alpha$- and $\beta$-rules until it becomes $E$-completed;[2]
3. if the resulting branch is neither completed nor closed then it selects a formula of type $\beta$ which is not yet fulfilled in the branch — if possible a $\beta$-formula which results from step 2 — then it applies PB with $\beta_1, \beta_1^C$ (or, equivalently $\beta_2, \beta_2^C$), and then it returns to step 1; otherwise it returns to step 1.

**Theorem 1.** *For a formula A, $A \equiv \top$ if either:*

1. *in one of the $\beta^C$-branches it generates there is a formula which appears twice, and one occurrence depends on $\beta_i^C, i \in \{1, 2\}$, and the other depends on $\beta$, or*
2. *each $\beta^C$-branch is closed and all the $\beta$-branches are $\top$-branches, or*
3. *each $\beta^C$-branch is a $\top$-branch.*

---

[2] For $\alpha$-formulas we do not duplicate components, i.e. if $\alpha$, and $\alpha_n$ ($n = 1, 2$) are in a branch, then we add to the branch only $\alpha_{3-n}$.

*Proof.* The proof follows from remarking that

|  tree for $\alpha$ | tree for $\beta$ $n = 1, 2$ |
|---|---|
| $\alpha$ | $\beta$ |
| $\alpha_1$ | $\beta_n^C \quad \beta_n$ |
| $\alpha_2$ | $\beta_{3-n}$ |

and

$$\alpha = \beta^C \qquad \alpha_1 = \beta_1^C \qquad \alpha_2 = \beta_2^C \qquad (5)$$
$$\beta = \alpha^C \qquad \beta_1 = \alpha_1^C \qquad \beta_2 = \alpha_2^C \qquad (6)$$

Thus a comparison of a $KE$-tree for $A^C$ [DM94] and a $KE^+$-tree for $A$ shows that whenever the $KE^+$-tree meets the conditions of the theorem the corresponding $KE$-tree is closed and vice versa.

**Definition 6.** *We define a function $v$ which maps each formula into a set of (atomic) formulas as follows: 1) if $A$ is atomic, then $v(A) = \{A\}$; 2) if $A$ is of type $\alpha$, then $v(A) = v(\alpha_1) \cup v(\alpha_2)$; 3) if $A$ is of type $\beta$, then $v(A) = v(\beta_n^C) \cup v(\beta_{3-n})$ or $v(A) = v(\beta_n)$, $n = 1, 2$. A set of (atomic) formulas $S$ is said to v-fulfill a formula $A$ iff $S = v(A)$.*

**Corollary 1.** *Two formulas $A, B$ are equivalent iff each set of (atomic) formulas which v-fulfils $A$ v-fulfils $B$.*

*Proof.* It is enough to notice that each set that $v$-fulfils a formula corresponds to a truth-value assignment for the formula, and two formulas are equivalent if they are satisfied by the same assignments.

### 4.5 Tableaux for $CL$s

Let us now analyse the tree for the formula 4 of Section 4.2 without using auxiliary trees.

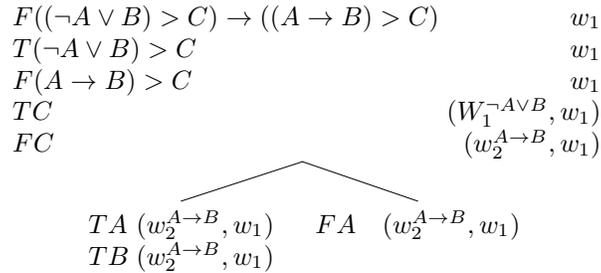

Obviously $\{TA, TB\}$ and $\{FA\}$ $v$-fulfill both $\neg A \vee B$ and $A \to B$, therefore $(W_1^{\neg A \vee B}, w_1)$ and $(w_2^{A \to B}, w_1)$ $\sigma_>$-unify, thus closing the tree.

We have still to examine the relations between formulas in the declarative part of $LS$-formulas and label formulas. Let us consider the following formula:

$$((A > B) \wedge (B > A)) \to ((A > C) \to (B > C)) \qquad (7)$$

This formula is the strong version of 3 and raises another problem of equivalence: the equivalence of two spheres. The corresponding tree is

| | |
|---|---|
| 1. $F((A > B) \wedge (B > A)) \to ((A > C) \to (B > C))$ | $w_1$ |
| 2. $T(A > B) \wedge (B > A)$ | $w_1$ |
| 3. $F(A > C) \to (B > C)$ | $w_1$ |
| 4. $TA > B$ | $w_1$ |
| 5. $TB > A$ | $w_1$ |
| 6. $TA > C$ | $w_1$ |
| 7. $FB > C$ | $w_1$ |
| 8. $FC$ | $(w_2^B, w_1)$ |
| 9. $TB$ | $(W_1^A, w_1)$ |
| 10. $TA$ | $(W_2^B, w_1)$ |
| 11. $TC$ | $(W_3^A, w_1)$ |

Notice that the formulas 9 and 10 in the above tree fulfill 4 and 5, from which the conditional equivalence of $A$ and $B$ in $w_1$ follows. We therefore need the following rule for detecting this fact,

$$\frac{TA, i^B \quad TB, j^A}{X, i^A}$$

if $h(i), h(j) \in \Phi_V$ and $(b(i), b(j))\sigma_L$.[3]

As regard the application of the method to conditional formulas allowing iteration of the conditional connective in the consequent, it suffices to remark that conditional formulas cannot occur as label formulass, and thus they do not play any rôle in $\sigma_>$-unifications.

## 5 Discussion and open problems

In this paper we have discussed a tableau methodology for building a computationally efficient tableau proof system for $CL$s. [GD88] and [Lam92] are already steps in this direction. Groeneboer and Delgrande [GD88] present a method for constructing Kripke models for $CL$ which generalizes Hughes and Cresswell's [HC68] method of semantic diagrams for the modal logic $S4.3$ to Delgrande's [Del87] conditional logic $NP$. This extension is made possible by the correspondence between $S4.3$ and $NP$ [Bou94]. However, to exploit this correspondence (via Kripke models for normal modal logics), we have to consider only the flat part of $NP$; conditional formulas allowing iterated conditionals in the consequent are excluded. Moreover, being an extension of Hughes and Cresswell [HC68]'s classic method of semantic diagrams for modal logics, Groeneboer and Delgrande's approach can be said to suffer of all the well-known computational drawbacks of the tableau method.

---
[3] See Section 3.1 for a similar consideration.

Lamarre [Lam92] takes a more direct approach by relying on $SOS$-models. However, his method requires the input formulas to be in "normal disjunctive form". This is a serious shortcoming, since normal forms are not generally available outside classical logic. On the contrary, the method we have proposed provides a uniform proof-theoretical treatment of $CL$s without normal-forming and, unlike [GD88]'s method, it does not require much declaratively expressed heuristic to improve efficiency in the various steps of the procedure.

In conclusion we wish to discuss some main sources of difficulties. According to [Alc96] a first problem with Lewis's systems is the world-relative character of their $SOS$ formal semantic models. As has been pointed out by Alchourrón "obligations are not world-relative but person-relative, and like defeasible conditionals, they require a person-relative construction". However, there should be no difficulties, in principle, to modify our approach so to adapt it to a person-relative construction of the logic of defeasible conditional obligation.

Another source of difficulties is that all Lewis's $CL$s for conditional obligation validate the deontically counterintuitive formula $A > A$ ("$A$ is obligatory given that $A$ is the case"). According to both [Alc96] and [Mak93] this difficulty can be bypassed by introducing some temporal ordering to make obligations future to their conditions. Unfortunately, this goes beyond Lewis's semantic equipment for $CL$s. Nevertheless we could introduce temporal distinctions in Lewis's semantics by holding the temporal interpretation of $SOS$-models (as *future* temporal systems of spheres) advanced in [Lew86], Section 5.2, together with their usual deontic interpretation and modifying accordingly the truth conditions for $>$ and our label formalism (e.g., by further labelling worlds with future time point or interval parameters). In this way, we could have $A$ true in a world $v$ after $u$ but false in some of the best $A$-worlds before $u$.

The most important source of difficulties has been scarcely discussed (see e.g. [Bou94,Mak93]). It comes from the fact that, as we have already said, most approaches to the logic of deontic, and in general defeasible, conditionality are restricted to "unnested" conditionals. In the AI & Law field this arises as a consequence of the fact that current non-monotonic or defeasible formalisms treat conditional obligation as a normal default (see e.g. [Hor94,Mak94,Pra93,TvT94]). However, perfectly understandable examples of nested conditionals occur both in ordinary language and in legal contexts. For example, judicial reasoning makes often use of the "weak" MP formula $(A \wedge (A > B)) > B$ to detach $B$ in circumstances in which $A$ and $A > B$ hold without known defeating conditions. Formulas with nested conditionals in the antecedent of this kind raise difficulties for traditional default formalisms. As we have said from the beginning, our approach suffers from the same weakness (though it is able to treat formulas with nested conditionals in the consequent).

A strictly related difficulty is that we can prove soundness and completeness of our method for the fragment $L_>^-$ of $L_>$ only by suspending the validity of the RCEA inference rule for normal $CL$s (from $A \equiv B$ to infer $(A > C) \equiv (B > C)$), or by restricting it to the propositional case (i.e. $A$, $B$ are conditional-free formulas). However, there are good reasons for rejecting RCEA in its usual, unrestricted formulation. Nute [Nut80], Ch. 2, has already argued for the inadequacy of RCEA for ordinary language conditionals. We believe that legal contexts offer further evidence of this inadequacy. For RCEA commits us to a truth-functional view of the "sameness of content" of two formulas $A$ and $B$ whereas, in contrast with classical propositional truths, conditional norms are states-of-affairs dependent (i.e., they have a non-null information content). Let us assume that a normative system can be conceived of as a logical theory, with the norms acting as its proper axioms[4]. On this view two (conditional) norms in it would be logically valid and thus truth-functionally equivalent. Let us further suppose that a normative system contains three norms $A > B$, $C > D$ and $(A > B) > E$. From RCEA we could then obtain $(C > D) > E$, in spite of the fact that the content (or the "field of application") of $A > B$ might be incompatible with, or inappropriate with respect to, that of $C > D$ and $E$.

---

[4] Of course, whether we take all norms in the system as its proper axioms or just a subset of them depends on whether the systems is taken to be finitely axiomatizable.